\documentstyle[pra,aps,epsf]{revtex}

\begin{document}
\def \beq {\begin{equation}}
\def \eeq {\end{equation}}
\def \eeqn {\end{equation}\noindent}
\def \bes {\begin{eqnarray}}
\def \ees {\end{eqnarray}}
\def \eesn {\end{eqnarray}\noindent}
\def \nn {\nonumber}
\def\mum{\,$\mu$m}

\draft
\thispagestyle{empty}

\title
{Perturbation approach to the Casimir force between
two bodies made of different real metals}

\author
{
B.~Geyer,  
G.L.~Klimchitskaya\footnote{On leave from North-West Polytechnical
University, \\
St.~Petersburg, Russia,  and Federal University of \\ 
Para\'{\i}ba,
Jo\~{a}o Pessoa, Brazil.},
V.M.~Mostepanenko\footnote{On leave from A.~Friedmann
Laboratory for Theoretical \\
Physics, St.Petersburg, 
Russia, and  Federal University  of \\ 
Para\'{\i}ba,
Jo\~{a}o Pessoa, Brazil.}}

\address
{Center of Theoretical Studies and Institute for Theoretical Physics,\\
Leipzig University, Augustusplatz 10/11, 04109, Leipzig, Germany }

\maketitle
\begin{abstract}
 The Casimir force acting between two test bodies made of different
metals is considered. The finiteness of the conductivity of the
metals is taken into account 
perturbatively up to the fourth order of the
relative penetration depths of electromagnetic zero-point oscillations
into the metals. The influence of nonzero temperature is computed
explicitly for separate orders of perturbation and found to be important
in the zeroth and first orders only. 
The configurations of two parallel plates and a sphere (spherical lens)
above a plate are considered made of $Au$ and $Cr$.
The obtained results can be used  
to take into account also the surface roughness. Thus, the total amount 
of the Casimir force between different metals with all correction factors 
is determined. This may be useful in various applications. 
\end{abstract}
\vskip 3mm

\section{Introduction}

Recently the Casimir effect attracted much attention as a macroscopic
quantum phenomenon caused by the existence of zero-point oscillations
of the electromagnetic field. Casimir \cite{1} first theoretically
proposed that the change of the zero-point oscillation spectrum in the
presence of metallic boundaries as compared to the case of empty space
leads to a finite force acting onto these boundaries. The Casimir
force can be considered as the relativistic limit of the van der Waals
force under the condition that the spatial separations between the surfaces
of the macrobodies are so large that the retardation effects become
essential. This was demonstrated at first qualitatively by Sparnaay
\cite{2}. During the last time a lot of precision experiments on
measuring the Casimir force have been performed \cite{3,4,5,6,7,8,9,10}.
The increasing interest in the Casimir effect is caused by the fact
that it found both fundamental as well as technological applications. Thus,
the precise measurements of the Casimir force and the extent of their 
agreement with theory have been used \cite{11,12,13,14} to set the
strongest constraints on hypothetical long-range forces predicted by
many extensions of the standard model of elementary particles. 
Concerning technological applications the first 
microelectromechanical machines were created being driven by the
Casimir force \cite{15,16}.

Increased precision and important applications of the Casimir force
measurements call for the elaboration of new computational methods
taking real experimental conditions into account. During the last years
different corrections to the ideal Casimir force were computed due to
surface roughness, finite conductivity of the boundary metal and 
nonzero temperature (see, e.g., papers 
\cite{17,18,18a,19,20,21,22,23,24,25,26,27} and review \cite{28}). Also
the combined effect of these influential factors was investigated for
the case of two boundary bodies being made of one and the same metal.
It was shown that at separations smaller than one micrometer the surface
roughness and the finite conductivity of the boundary metal can contribute
up to several tens percent of the ideal Casimir force. At the same time,
at separations of order of several micrometers the temperature corrections 
can achieve the value of the main contribution and even become larger.
In a transition range of separations all the above corrections play an
important role and their combined effect must be considered. However,
the case of boundary bodies made of different metals was not
investigated up to now.

Here, we present a perturbative approach to the calculation of the
Casimir force acting between two bodies made of different real metals.
We start from the famous Lifshitz formula \cite{29} and describe metals
in the framework of the plasma model. Both the configurations of two plane
parallel plates and a sphere (spherical lens) above a plate are
considered. The combined effect of finite conductivity and nonzero
temperature is found on the basis of a perturbation expansion in powers
of the relative penetration depth of electromagnetic zero-point
oscillations into the metals under consideration. The coefficients of this
expansion up to the fourth order are calculated explicitly.
The temperature effect is shown to be essential only in the 
zeroth- and first-order terms. 
The obtained results are generalizations of those which previously have been 
obtained in \cite{21,23,25} for the case of boundary bodies
made of one and the same metal. The case of different metals considered 
here is of especial importance for the nanotechnology where one of the 
plates (playing the role of an active element) and the underlying substrate
are typically made of different materials. The perturbation formulas 
given below are very simple in its application and open the possibility to
compute the combined effect of finite conductivity and nonzero
temperature with an error not larger than 1--2\% within a wide separation
range which is usually adequate for any practical purposes. In doing so 
one avoids labour-intensive numerical computations based on the use of optical
tabulated data for the complex refraction index (compare with \cite{19,20}
where such computations were performed for two bodies made of one and
the same metal). As an example, the test bodies covered by $Au$ and $Cr$ 
layers are considered. 
The obtained formulas can simply be modified to take
into account the surface roughness (see, e.g., \cite{25,28} for the special
averaging procedure). Thus, they can be used for a complete description of
the Casimir force with all essential corrections.

The paper is organized as follows. In Sec.~II the perturbation
expansion for the Casimir force at zero temperature acting between plates 
made of different metals is derived. Sec.~III is devoted to the
consideration of temperature corrections in configurations of two
different plates. The configuration of a sphere (spherical lens) above
a plate made of different metals is considered in Sec.~IV.
Sec.~V contains conclusions and discussion.

\section{Perturbation expansion for two parallel plates made of different
metals at zero temperature}

We consider first the configuration of two metallic semispaces
marked by an index $ss$ and described by the dielectric permittivities
$\varepsilon_1(\omega)$ and $\varepsilon_2(\omega)$, respectively.
Let these semispaces be separated by a plane parallel gap of width $a$.
The Casimir force between two different metals at zero temperature
is given by the Lifshitz formula \cite{29,30,31}
\beq
F_{ss}^{(\delta,0)}(a)=
-\frac{\hbar c}{32\pi^2 a^4}
\int_{0}^{\infty}x^3\,dx
\int_{1}^{\infty}\frac{dp}{p^2}
\left[X_1(p,x)
+X_2(p,x)\right],
\label{1}
\eeqn
where
\beq
X_1(p,x)=\left[\frac{(s_1+p\varepsilon_1)(s_2+
p\varepsilon_2)}{(s_1-p\varepsilon_1)(s_2-
p\varepsilon_2)}e^{x}-1\right]^{-1},
\quad
X_2(p,x)=\left[\frac{(s_1+p)(s_2+p)}{(s_1-p)(s_2-p)}
e^{x}-1\right]^{-1}.
\label{2}
\eeqn
The quantities $s_k$ ($k=1,\,2$) are given by
$s_k=\sqrt{\varepsilon_k-1+p^2}$, and the dielectric permittivities
are computed on the imaginary frequency axis 
$\varepsilon_k\equiv\varepsilon_k(i\xi)=
\varepsilon_k[icx/(2pa)]$. The upper index $\delta$ in (\ref{1})
marks the account of the effect of finite conductivity, and the
second upper index 0 is the value of temperature.  

For separation ranges of practical interest, namely from a few tens of a 
micrometer to ten micrometers, relaxation processes can be neglected and the 
dielectric permittivities of the metals are given by the free electron
plasma model,
\beq
\varepsilon_k(i\xi)=1+\frac{\omega_{pk}^2}{\xi^2},
\label{3}
\eeqn
where $\omega_{pk}$ are the plasma frequencies of the metals under
consideration. The perturbative approach for calculating the effect
of finite conductivity is based on the use of small parameters,
\beq
\alpha_k=\frac{\xi}{\omega_{pk}}=
\frac{\delta_k}{a}\,\frac{x}{2p},
\label{4}
\eeqn
where $\delta_k=\lambda_{pk}/(2\pi)$ are the effective penetration depths
of the electromagnetic zero-point oscillations into the metals and
$\lambda_{pk}=2\pi c/\omega_{pk}$ are the plasma wavelengths. For the
case of plates made of one and the same metal this approach was used
in \cite{30,32,33} (up to the first order), in \cite{34} (up to the second
order) and in \cite{21,35} up to the fourth and the sixth orders,
respectively (see also the detailed explanations in the monographs
\cite{36,37,38}). The obtained results were compared with the numerical
computations using the tabulated data for the complex refraction
index and were found to be in agreement with an error of only 1--2\% at
all separation distances larger than the plasma wavelength \cite{20,28}.
Here, we apply this approach to the case of different metals up to the
fourth perturbative order which is sufficient for practical purposes.

The quantities $\varepsilon_k$ and $s_k$ entering Eq.~(\ref{2}) can be
represented in terms of the small parameters (\ref{4}) as
\beq
\varepsilon_k(i\xi)=1+\frac{1}{\alpha_k^2},
\qquad
s_k=\sqrt{p^2+\frac{1}{\alpha_k^2}}.
\label{5}
\eeqn
Expanding $X_1$ from Eq.~(\ref{2}) up to the fourth power in $\alpha_k$
one obtains
\bes
&&
X_1=\frac{1}{e^x-1}\left\{1-2\frac{A}{p}(\alpha_1+\alpha_2)+
2\frac{A}{p^2}(2A-1)(\alpha_1+\alpha_2)^2\right.
-\frac{A}{p^3}(2-8A+8A^2-2p^2+p^4)(\alpha_1^3+\alpha_2^3)
\nn \\
&&\phantom{aaa}
-4\frac{A}{p^3}(1-6A+6A^2)(\alpha_1^2\alpha_2+\alpha_1\alpha_2^2)
+2\frac{A(2A-1)}{p^4}\left[(2A-1)^2-2p^2+p^4\right](\alpha_1^4+\alpha_2^4)
\label{6} \\
&&\phantom{aaa}
+2\frac{A(2A-1)}{p^4}(2-16A+16A^2-2p^2+p^4)(\alpha_1^3
\alpha_2+\alpha_1\alpha_2^3)
\left.
+4\frac{A(2A-1)}{p^4}(1-12A+12A^2)\alpha_1^2
\alpha_2^2\right\},
\nn
\eesn
where $A\equiv e^x/(e^x-1)$.

In the same way the expansion of $X_2$ is
\bes
&&
X_2=\frac{1}{e^x-1}\left[1-2Ap(\alpha_1+\alpha_2)+
2Ap^2(2A-1)(\alpha_1+\alpha_2)^2\right.
-Ap^3(1-8A+8A^2)(\alpha_1^3+\alpha_2^3)
\nn \\
&&\phantom{aaa}
-4Ap^3(1-6A+6A^2)(\alpha_1^2\alpha_2+\alpha_1\alpha_2^2)
+8A^2p^4(1-3A+3A^2)(\alpha_1^4+\alpha_2^4)
\label{7} \\
&&\phantom{aaa}
-2Ap^4(1-18A+48A^2-32A^3)(\alpha_1^3
\alpha_2+\alpha_1\alpha_2^3)
\left.
-4Ap^4(1-14A+36A^2-24A^3)\alpha_1^2
\alpha_2^2\right].
\nn
\eesn
Substituting expressions (\ref{6}) and (\ref{7}) into Eq.~(\ref{1}) 
and performing
the integrations with respect to $p$ and $x$ one finally obtains
\bes
&&
F_{ss}^{(\delta,0)}(a)=F_{ss}^{(0,0)}(a)\left\{
\vphantom{\left[1-\frac{326\pi^2}{3675}(1-3\kappa)\right]}
1-\frac{16}{3}\,\frac{\delta}{a}+24\frac{\delta^2}{a^2}
\right.
\nn \\
&&\phantom{aaa}
-\frac{640}{7}\,\frac{\delta^3}{a^3}
\left[1-\frac{2\pi^2}{105}(1-3\kappa)\right]
+\left.
\frac{2800}{9}\,\frac{\delta^4}{a^4}
\left[1-\frac{326\pi^2}{3675}(1-3\kappa)\right]\right\},
\label{8}
\eesn
where
\beq
\delta\equiv\frac{\delta_1+\delta_2}{2},
\qquad
\kappa\equiv\frac{\delta_1\delta_2}{(\delta_1+\delta_2)^2},
\label{9}
\eeqn
and $F_{ss}^{(0,0)}(a)=\pi^2\hbar c/(240a^4)$ is the ideal Casimir
force per unit area of plates made of perfect metal. If 
$\delta_1=\delta_2=\delta_0$, i.e., when the plates are made of one and
the same metal, Eq.~(\ref{8}) coincides with the result obtained
earlier in Refs.~\cite{21,35}.

\section{Two parallel plates made of different metals at nonzero
temperature}

Now let us consider the case of nonzero temperature. The Lifshitz formula at
$T\neq 0$ is obtained from Eq.~(\ref{1}) by changing the integration
with respect to $x$ into a summation over the discrete Matsubara
frequencies 
$\xi_l=cx_l/(2ap)=2\pi l k_BT/\hbar$, where $l=0,\pm1,\pm2,...\,$ and 
$k_B$ is the Boltzmann constant. It is convenient also to introduce
a new variable $k_{\bot}=\xi_l\sqrt{p^2-1}/c$ \cite{28}. The result is
\beq
F_{ss}^{(\delta,T)}(a)=-\frac{k_BT}{2\pi}
\sum\limits_{l=-\infty}^{\infty}
\int_{0}^{\infty}
k_{\bot}\,dk_{\bot}
\sqrt{\frac{\xi_l^2}{c^2}+k_{\bot}^2}
\left[X_1(k_{\bot},\xi_l)+X_1(k_{\bot},\xi_l)\right].
\label{10}
\eeqn
Using the Poisson summation formula this expression can be represented
as the sum of the zero-temperature result (\ref{1}) and a temperature
correction.

It can be easily checked that the temperature corrections of the first
expansion coefficients of the perturbation result (\ref{8}) following
from (\ref{1}) are independent on the materials, i.e., they are the same 
for plates made of one and the same metal and of different metals. 
In the framework of the plasma model these corrections can be
calculated analytically in a closed form without using any perturbation
expansion (see Refs.\cite{39,22,25} where the corrections to the 
zeroth-, first-, and second-order coefficients, respectively, were found
for the plates made of one and the same metal). 
It was proved in \cite{24,25,27} that the plasma model is well adapted
to the Lifshitz formula (\ref{10}) at nonzero temperature and that it avoids
all problems and contradictions arising in the case of Drude
dielectric function. As it was shown in 
Ref.\cite{25}, in the temperature range from 0\,K to 1000\,K the
temperature corrections to the expansion coefficients starting from
the second-order one are not essential. The reason is that at small
surface separations the temperature effect itself is negligible
whereas at large separations the contribution of all the terms of
order $(\delta/a)^k$ with $k\geq 2$ is smaller than 1\%. This opens the
opportunity to modify Eq.~(\ref{8}) by the use of temperature
corrections, calculated in \cite{22,25,39} in order to obtain the
approximate expression for the Casimir force acting between different
metals with account of both finite conductivity and nonzero temperature.
The final result can be represented in the form
\bes
&&
F_{ss}^{(\delta,T)}(a)=F_{ss}^{(0,0)}(a)\left\{
1+\frac{30}{\pi^4}
\sum\limits_{n=1}^{\infty}
\left[\frac{1}{(nt)^4}-\frac{\pi^3}{nt}\,
\frac{\coth(\pi nt)}{\sinh^2(\pi nt)}\right]
\right.
-2\frac{\delta}{a}\left[\frac{8}{3}-\frac{15}{\pi}
\sum\limits_{n=1}^{\infty}
\frac{1}{nt\sinh^2(\pi nt)}
\right.
\nn \\
&&\phantom{aa}
\times\left(\frac{1}{(\pi nt)^2}
\sinh(\pi nt)\cosh(\pi nt)
\left.\vphantom{\frac{1}{nt\sinh^2(\pi nt)}}
+4\coth(\pi nt)+2\pi nt-6\pi nt\coth^2(\pi nt)+
\frac{1}{\pi nt}\right)
\vphantom{\sum\limits_{n=1}^{\infty}
\frac{1}{nt\sinh^2(\pi nt)}\left(\frac{1}{(\pi nt)^2}\right)}
\right]
\label{11} \\
&& \phantom{aa}
+24\frac{\delta^2}{a^2}
-\frac{640}{7}\,\frac{\delta^3}{a^3}
\left[1-\frac{2\pi^2}{105}(1-3\kappa)\right]
+\left.
\frac{2800}{9}\,\frac{\delta^4}{a^4}
\left[1-\frac{326\pi^2}{3675}(1-3\kappa)\right]
\vphantom{\sum\limits_{n=1}^{\infty}
\left[\frac{1}{(nt)}^4-\frac{\pi^3}{nt}\,
\frac{\coth(\pi nt)}{\sinh^2(\pi nt)}\right]}
\right\},
\nn
\eesn
where $t\equiv T_{eff}/T$, and $k_BT_{eff}\equiv\hbar c/(2a)$.

One can easily find the asymptotic behavior of Eq.~(\ref{11}) at low 
($T\ll T_{eff}$) and high ($T\gg T_{eff}$) temperatures (which also
means small, respectively, large separations when taking into account
the definition of $T_{eff}$). At low temperatures ($t\gg 1$)
it holds
\bes
&&
F_{ss}^{(\delta,T)}(a)\approx F_{ss}^{(0,0)}(a)\left\{
\vphantom{\frac{2800}{9}\,\frac{\delta^4}{a^4}
\left[1-\frac{326\pi^2}{3675}(1-3\kappa)\right]}
1+\frac{1}{3t^4}-
2\frac{\delta}{a}\left[\frac{8}{3}-\frac{15}{\pi^3t^3}\zeta(3)\right]
\right.
\nn \\
&&\phantom{aaa}
+24\frac{\delta^2}{a^2}-\frac{640}{7}\,\frac{\delta^3}{a^3}
\left[1-\frac{2\pi^2}{105}(1-3\kappa)\right]
+\left.
\frac{2800}{9}\,\frac{\delta^4}{a^4}
\left[1-\frac{326\pi^2}{3675}(1-3\kappa)\right]
\right\},
\label{12}
\eesn
where $\zeta(z)$ is the Riemann zeta function.
At high temperatures ($t\ll 1$) the result is given by
\beq
F_{ss}^{(\delta,T)}(a)\approx F_{ss}^{(0,0)}(a)
\frac{30\zeta(3)}{\pi^3t}\left(1-3\frac{\delta}{a}\right).
\label{13}
\eeqn
All the results (\ref{11})--(\ref{13}) take into account that the metals of
both plates are different.

By way of example, in Table I some numerical data obtained by
Eq.~(\ref{11}) are presented for the pairs of plates $Au-Au$, $Au-Cr$ and
$Cr-Cr$. Note that both $Au$- and $Cr$-covered test bodies are widely
used in the measurements of the Casimir force (see, e.g., 
\cite{3,7,8,9,10,15,16,40}).
For $Au$ the value of the plasma wavelength $\lambda_{p1}=136\,$nm
was used \cite{19} and for $Cr$ $\lambda_{p2}=314\,$nm \cite{40}.
The separations range 0.35--10\,$\mu$m is covered including both
the cases of low and high temperatures. The smallest separation
0.35\,$\mu$m is chosen to be larger than both plasma wavelengths in
order to respect the application range of the four-order perturbation
expansion of Eq.~(\ref{11}). In Table I the ratio of the Casimir force
acting between real metals at zero and room temperatures relative to the ideal
value (i.e. to a force between perfect metals at zero temperature) is
given. In the last column the absolute values of the ideal Casimir
force in units of force per unit area are presented. As is seen from
Table I, the effect of finite conductivity is especially important at
the smallest separations. The results for the pair of different metals 
($Au-Cr$) differ significantly from both the cases $Au-Au$ and $Cr-Cr$. 
At small separations the temperature effect is negligible. With an
increase of the separation distance also the temperature effect increases in 
all cases, and for separations larger than 3\,$\mu$m it becomes larger
than the ideal Casimir force. However, even at largest separations
under consideration the effects of finite conductivity influence the
value of the temperature force. Note that the asymptotics of low
temperatures (\ref{12}) is applicable at separations smaller than
2\,$\mu$m and the asymptotics of high temperatures (\ref{13}) works
good starting from 4\,$\mu$m.

\section{Configuration of a sphere above a plate made of different metals}

The configuration of two plane parallel plates was used only in two
experiments \cite{2,9}. More often the configuration of a sphere (spherical
lens) above a plate was employed \cite{3,4,5,6,7,8,10,15,16,40}.
By this reason it is expedient to modify the obtained results for this 
configuration. This can be achieved by the application of the Proximity Force 
Theorem \cite{41}. According to this theorem the Casimir force acting
between a semispace and a lens $F_{sl}(a)=2\pi R E_{ss}(a)$, where
$R$ is the lens (sphere) radius, and $E_{ss}(a)$ is the energy per unit area
of two parallel plates which is related 
to the force of Eqs.~(\ref{1}) and (\ref{8})
by the equality $F_{ss}(a)=-\partial E_{ss}(a)/\partial a$. Although
the Proximity Force Theorem is an approximation, its accuracy is very
high (it leads to an error of order $a/R$ which is much smaller than 1\%
for configurations used in the experiments \cite{42,43}).

Applying the Proximity Force Theorem to Eq.~(\ref{8}) one obtains the 
Casimir force acting between a plate and a lens (sphere) made of
different real metals at zero temperature
\bes
&&
F_{sl}^{(\delta,0)}(a)=F_{sl}^{(0,0)}(a)\left\{
\vphantom{\left[1-\frac{326\pi^2}{3675}(1-3\kappa)\right]}
1-4\frac{\delta}{a}+\frac{72}{5}\,\frac{\delta^2}{a^2}
\right.
\nn \\
&&\phantom{aaa}
-\frac{320}{7}\,\frac{\delta^3}{a^3}
\left[1-\frac{2\pi^2}{105}(1-3\kappa)\right]
+\left.
\frac{400}{3}\,\frac{\delta^4}{a^4}
\left[1-\frac{326\pi^2}{3675}(1-3\kappa)\right]\right\},
\label{14}
\eesn
where $F_{sl}^{(0,0)}(a)=-\pi^3\hbar cR/(360a^3)$ is the ideal Casimir
force. If one puts $\delta_1=\delta_2=\delta_0$ then Eq.~(\ref{14}) coincides
with the earlier result for test bodies made of one and the same
metal \cite{21,35}.

In the same way as it was done for the two parallel plates, Eq.~(\ref{14})
can be generalized to take into account nonzero temperature in the
range from 0\,K to 1000\,K. Using the results of Ref.~\cite{25},
one obtains
\bes
&&
F_{sl}^{(\delta,T)}(a)=F_{sl}^{(0,0)}(a)\left\{
1+\frac{90}{\pi^4}
\sum\limits_{n=1}^{\infty}\left[\frac{\pi}{2(nt)^3}\coth(\pi nt)
\right.\right.
\left.
-\frac{1}{(nt)^4}+\frac{\pi^2}{2(nt)^2}\,\frac{1}{\sinh^2(\pi nt)}
\right]
\label{15} \\
&&\phantom{aaa}
-2\frac{\delta}{a}\left[2-\frac{45}{\pi^4}
\sum\limits_{n=1}^{\infty}\left(\frac{\pi}{(nt)^3}\coth(\pi nt)
-\frac{4}{(nt)^4}\right.
\right.
\left.\left.
+\frac{\pi^2}{(nt)^2}\,\frac{1}{\sinh^2(\pi nt)}
+\frac{2\pi^3}{nt}\,\frac{\coth(\pi nt)}{\sinh^2(\pi nt)}
\right)
\vphantom{\sum\limits_{n=1}^{\infty}\left(\frac{\pi}{(nt)^3}\right)}
\right]
\nn \\
&&\phantom{aaa}
+\frac{72}{5}\,\frac{\delta^2}{a^2}-\frac{320}{7}\,\frac{\delta^3}{a^3}
\left[1-\frac{2\pi^2}{105}(1-3\kappa)\right]
+\left.
\frac{400}{3}\,\frac{\delta^4}{a^4}
\left[1-\frac{326\pi^2}{3675}(1-3\kappa)\right]
\vphantom{\sum\limits_{n=1}^{\infty}\left[\frac{\pi}{2(nt)^3}\right]}
\right\}.
\nn
\ees

The asymptotic behavior of Eq.~(\ref{15}) at low temperatures
(separations) is
\bes
&&
F_{sl}^{(\delta,T)}(a)\approx F_{sl}^{(0,0)}(a)\left\{
\vphantom{\frac{400}{3}\,\frac{\delta^4}{a^4}
\left[1-\frac{326\pi^2}{3675}(1-3\kappa)\right]}
1+\frac{45\zeta(3)}{\pi^3t^3}-\frac{1}{t^4}
\right.
-2\frac{\delta}{a}\left[2-\frac{45\zeta(3)}{\pi^3t^3}+\frac{2}{t^4}
\right]
\label{16} \\
&&\phantom{aaa}
+\frac{72}{5}\,\frac{\delta^2}{a^2}-\frac{320}{7}\,\frac{\delta^3}{a^3}
\left[1-\frac{2\pi^2}{105}(1-3\kappa)\right]
+\left.
\frac{400}{3}\,\frac{\delta^4}{a^4}
\left[1-\frac{326\pi^2}{3675}(1-3\kappa)\right]
\right\}.
\nn
\ees

In the opposite case of high temperatures (large separations) the
asymptotic behavior is
\beq
F_{sl}^{(\delta,T)}(a)\approx F_{sl}^{(0,0)}(a)\frac{45\zeta(3)}{\pi^3t}
\left(1-2\frac{\delta}{a}\right).
\label{17}
\eeq

As an example, in Table II the numerical results obtained by 
Eq.~(\ref{15}) are presented for the case of a plate and a sphere made of
$Au-Au$, $Au-Cr$ and $Cr-Cr$. Table II is organized in the same manner as 
Table I -- only the configuration of the test bodies is different.
The last column of Table II contains the values of the ideal Casimir
force for a sphere of radius $R=1\,$mm.
The data demonstrate almost the same behavior with the increase of
the separation as in the case of two plane parallel plates. It is seen,
however, that the effect of nonzero temperature becomes noticeable
at smaller separations. The asymptotics of low temperatures (\ref{16})
works good at separations smaller than 2\,$\mu$m and the asymptotics
of high temperatures (\ref{17}) is applicable for $a>4\,\mu$m.

\section{Conclusions and discussion}

To conclude, we have developed a perturbative approach to the calculation 
of the Casimir force acting between two parallel plates or a sphere 
(spherical lens) above a plate made of different real metals.
The coefficients of the perturbation expansion in powers of two small
parameters were found up to the fourth order. These parameters have
the meaning of the effective penetration depth of electromagnetic
zero-point oscillations into both metals. The effect of nonzero
temperature was taken into account explicitly in the coefficients of
perturbation expansions of zeroth and first orders. The temperature
dependence of the higher order expansion coefficients is negligible
in the temperature range from 0\,K to 1000\,K and thereby it is of no
practical interest. The asymptotic behavior of the explicit temperature
dependences at low and high temperatures is also given.

The obtained formulas are simple in application and give the possibility
to calculate the Casimir force with account of both finite conductivity
and nonzero temperature between the test bodies made of different metals.
They can be applied in a wide range of separations and temperatures
quite sufficient for all practical purposes. The error of the results
obtained in such a way is of only 1--2\% and is caused in fact by the
error in the values of plasma wavelengths. The much more complicated
alternative approach using the optical tabulated data for the complex
refraction index and the exact Lifshitz formula does not lead to more
exact results because of the errors in optical data and the necessity to use
some interpolation and extrapolation procedures \cite{19}.
It is notable also that the above perturbative approach is very convenient
to take into account the surface roughness. This can be done by 
averaging of the obtained results over all possible separation
distances and it leads to a perfect agreement between experiment and theory
\cite{10,17}.
Thus, the suggested perturbative approach presents a complete description of 
the Casimir force acting between different metals with all important
corrections and can be used in various applications of the Casimir effect.

\section*{Acknowledgments}

G.L.K. and V.M.M. are indebted to the Center of Theoretical Studies and
the Institute for Theoretical
Physics, Leipzig University, for kind hospitality. Their work was
supported by the Saxonian Ministry of Science and Fine Arts (Germany)
and by CNPq (Brazil).

\widetext
\begin{table}[h]
\caption{The relative Casimir force between two parallel plates
with account of finite conductivity and temperature corrections
versus separation for different pairs of metals}
\begin{tabular}{cccccccc}
Separation &\multicolumn{2}{c}{$F_{ss}^{(\delta,T)}/F_{ss}^{(0,0)}$
for $Au-Au$}&\multicolumn{2}{c}{$F_{ss}^{(\delta,T)}/F_{ss}^{(0,0)}$
for $Au-Cr$}&\multicolumn{2}{c}{$F_{ss}^{(\delta,T)}/F_{ss}^{(0,0)}$
for $Cr-Cr$}& $F_{ss}^{(0,0)}(a)$\\
$a$($\mu$m)& T=0\,K& T=300\,K&T=0\,K& T=300\,K&T=0\,K& T=300\,K&
(nN/mm${}^2$)\\
0.35 & 0.745 &0.745 & 0.637 & 0.637 & 0.575 & 0.575 & 86.6 \\
0.4 & 0.770 & 0.770 & 0.667 & 0.667 & 0.597 & 0.597 & 50.8 \\
0.6 & 0.835 & 0.835 & 0.752 & 0.752 & 0.684 & 0.684 & 10.0 \\
0.8 & 0.872 & 0.873 & 0.803 & 0.804 & 0.743 & 0.744 & 3.17 \\
1 & 0.895 & 0.897 & 0.836 & 0.838 & 0.784 & 0.786 & 1.30 \\
3 & 0.963 & 1.083 & 0.940 & 1.062 & 0.917 & 1.042 & 1.60$\times 10^{-2}$ \\
5 & 0.977 & 1.531 & 0.963 & 1.518 & 0.949 & 1.505 & 2.08$\times 10^{-3}$ \\
7 & 0.984 & 2.116 & 0.973 & 2.104 & 0.963 & 2.091 & 5.41$\times 10^{-4}$ \\
10 & 0.988 & 3.027 & 0.981 & 3.015 & 0.974 & 3.002 & 3.05$\times 10^{-4}$
\end{tabular}
\end{table}
\begin{table}[h]
\caption{The relative Casimir force between a lens and a plate
with account of finite conductivity and temperature corrections
versus separation for different pairs of metals}
\begin{tabular}{cccccccc}
Separation &\multicolumn{2}{c}{$F_{sl}^{(\delta,T)}/F_{sl}^{(0,0)}$
for $Au-Au$}&\multicolumn{2}{c}{$F_{sl}^{(\delta,T)}/F_{sl}^{(0,0)}$
for $Au-Cr$}&\multicolumn{2}{c}{$F_{sl}^{(\delta,T)}/F_{sl}^{(0,0)}$
for $Cr-Cr$}& $F_{sl}^{(0,0)}(a)$\\
$a$($\mu$m)& T=0\,K& T=300\,K&T=0\,K& T=300\,K&T=0\,K& T=300\,K&
(nN)\\
0.35 & 0.799 &0.800 & 0.706 & 0.707 & 0.639 & 0.640 & 6.45$\times 10^{-2}$ \\
0.4 & 0.820 & 0.822 & 0.732 & 0.735 & 0.665 & 0.668 & 4.25$\times 10^{-2}$ \\
0.6 & 0.872 & 0.879 & 0.805 & 0.811 & 0.746 & 0.754 & 1.26$\times 10^{-2}$ \\
0.8 & 0.902 & 0.916 & 0.846 & 0.862 & 0.797 & 0.813 & 5.32$\times 10^{-3}$ \\
1 & 0.920 & 0.947 & 0.873 & 0.902 & 0.831 & 0.860 & 2.72$\times 10^{-3}$ \\
3 & 0.972 & 1.443 & 0.954 & 1.427 & 0.937 & 1.411 & 1.01$\times 10^{-4}$ \\
5 & 0.983 & 2.275 & 0.972 & 2.262 & 0.961 & 2.249 & 2.18$\times 10^{-5}$ \\
7 & 0.988 & 3.181 & 0.980 & 3.168 & 0.972 & 3.155 & 7.93$\times 10^{-6}$ \\
10 & 0.991 & 4.551 & 0.986 & 4.538 & 0.980 & 4.526 & 2.72$\times 10^{-6}$
\end{tabular}
\end{table}
\end{document}